# EVALUATION OF COMPUTABILITY CRITERIONS FOR RUNTIME WEB SERVICE INTEGRATION

Thirumaran.M[1],  Dhavachelvan.P[2], Aranganayagi.G[3] and S.Abarna[4]

[1,3,4] Department of Computer Science and Engineering, Pondicherry Engg College, India.

[2] Department of Computer Science and Engineering, Pondicherry University, India.

## ABSTRACT

*Today's competitive environment drives the enterprises to extend their focus and collaborate with their business partners to carry out the necessities. Tight coordination among business partners assists to share and integrate the service logic globally. But integrating service logics across diverse enterprises leads to exponential problem which stipulates developers to comprehend the whole service and must resolve suitable method to integrate the services. It is complex and time-consuming task. So the present focus is to have a mechanized system to  analyze the Business logics and convey the  proper mode to integrate them. There is no standard model to undertake these issues and one such a framework proposed in this paper examines the Business logics individually and suggests proper structure to integrate them. One of the innovative concepts of proposed model is Property Evaluation System which scrutinizes the service logics and generates Business Logic Property Schema (BLPS) for the required services. BLPS holds necessary information to recognize the correct structure for integrating the service logics. At the time of integration, System consumes this BLPS schema and suggests the feasible ways to integrate the service logics. Also if the service logics are attempted to integrate in invalid structure or attempted to violate accessibility levels, system will throw exception with necessary information. This helps developers to ascertain the efficient structure to integrate the services with least effort.*

## 1. INTRODUCTION

With the trend in economic globalization and enormous development in information technology, the demand for information and logic sharing has become more serious which urges the companies to collaborate closely with their business partners to gain access to needed information and business logic. Over the past decade, the companies have been using various technologies and products in an attempt to support collaboration. These solutions vary from basic point-to-point connection approach such as EDI, expensive ERP systems such as Rossetanet, ebXML, etc.  The current technologies semi-automatically integrate the services and it needs manual intervention in number of areas. It requires developers to analyze the service to identify possible way for integration. It is a complex task which needs developers to understand both service and identify better way for integration. Also the present technologies does not consider how to composite of services and how to describe the service contracts. We proposed Business Logic model to face these brutal challenge and complexities.

The proposed model enables the automation of service integration by coordinating sequences of tasks and supports sophisticated exception management. The proposed Business Logic Model uses property evaluator method to evaluate the service to ascertain correct structure for integration. It analyses at which level service fulfills particular property in functionality level and also as per contract, accordingly develops flow diagram as it reflects property evaluation outcome. Then BLP (Business Logic Property) schema is generated from this diagram holding necessary information for integration. While integration, System utilizes this BLP schema to identify proper structure for integration and to spot various actions can be carried out with the service. With this flocked information from BLP schema, it integrates the service automatically. If services are integrated as violating contract or with invalid structure, the system will throw exception with necessary information. End-to-end security is provided by annotating service descriptions with security objectives used to generate convenient Quality of Protection Agreements between partners. Conversely, agreements are processed by a dedicated matching module with respect to security requirements stated by the SLA. In addition to this, we need a mechanism to monitor the resource while sharing to adapt the modifications made by the developers. Source control Management tracks the modification and facilitates impact analysis between the existing and modified services that ensures computability criteria. The source control management system allows us to see the historical background behind the changes made to the business logic of the web services. This helps the developers to see where the changes have been progressively made and include or remove the change as per the need. Thus this would be a powerful and easiest model for developers to integrate the services. Here we demonstrated service integration with BLP schema generation for banking application using Netbeans IDE.

## 2. RELATED WORKS

In this section, we discuss various research work and different solutions exist in the market for service integration. Zuoren Jiang proposed a model called 'Multi-layer Structure for Dynamic Service Integration (MSFDSI)' in SOA which adds authorized institution and a service integration & analysis adapter to achieve the service authorization, service analysis and dynamic service integration. Service integration & analysis adapter analyses and search the service that can meet the service requestor's requests according to service contracts stated by authorized institution [1]. W.J.Yan proposed B2B integration approach for SME which provides a feasible and cost-effective B2Bi solution for SMEs by leveraging the characteristics of Web Services. It utilizes pull and push mechanisms for effective information exchange and sharing between trading partners. This approach has been incorporated in a B2Bi Gateway which enables SMEs to participate in business-to-business collaboration by making use of Web Services [2]. Liyi Zhang proposed a model called WSMX (Web Service Modeling execution), a software system that enables the creation and execution of Semantic Web Services based on the Web Service Modeling Ontology (WSMO) for enterprise application integration. It improves Service discovery, simplifies change management and supports semi-automatic service composition and enhanced interoperability between services [3]. Thomas Haselwanter presented a model based on the WSMX was build to tackle heterogeneities in RosettaNet messages by using the axiomatised knowledge and rules. It supports communication between partners, data and process mediation using WSMX integration middleware[4]. Jianwei Yin proposed an ESB framework for large scale Service Integration, JTangSynergy adopts several mechanisms for providing effective and efficient dependability. It enables automated recovery from component failures and robust execution of composite services by checking service compatibility [5].

Gulnoza Ziyaeva proposed framework to enable the content-based intelligent routing path construction and message routing in ESB which defines the routing tables and mechanisms of message routings and facilitate the service selection based on message content [6]. Soo Ho Chang proposed a framework for dynamic composition on Enterprise Service Bus which consists of four elements; Invocation Listener, Service Router, Service Discoverer, and Interface Adapter. This framework enables the runtime discovery and composition of published services without altering the client side applications [7]. Liu Ying presents a unified service composition framework to support business level service composition. An intelligent service composer based on this unified service composition framework is developed to enable business level service composition by business people under the help of some advanced technologies, including intelligent service components searching, automatic service compliance checking, and template-based service adaptation [8]. In addition, Companies use different solutions exist in the market for Business to business application framework, including EDI, RosettaNet, ebXML etc. EDI: A seminal event in B2B evolution was the development of electronic data interchange (EDI), whereby trading partners established standard formats for the exchange of electronic documents to facilitate electronic transactions. Trading partnerships between two firms using EDI are well defined and is used for automated replenishment and efficient supply chains[9]. RosettaNet: The RosettaNet consortium develops XML-based business standards for supply chain management in the information technology and electronic component industries. It defines the business processes and provides the technical specifications for data interchange. RosettaNet standards comprise Dictionary, RNIF (RosettaNet Implementation Framework) and PIP (Partner Interface Process)[10][11]. ebXML: The electronic business XML (ebXML) provides a complete framework for setting up B2B collaborations. It is a set of documents, with several prototype completed, enabling businesses of any size to do business electronically with anyone else. The ebXML specifications cover almost the entire B2B collaboration process: collaboration Protocol Profile (CPP), Collaboration Protocol Agreement (CPA), Business Process Specification Schemas (BPSS), Messaging, Registry/repository and a core Component [12]. Above works paves way to semi-automatically integrate the services across enterprise. But still there is no mechanism to monitor the services while sharing and to routinely guide the developers to integrate according to SLA. Here we demonstrated service integration with BLP schema generation for banking application using Netbeans IDE.

## 3 BUSINESS LOGIC MODEL

Figure 1 depicts detailed architecture and illustrates how enterprises integrate their services dynamically. Let Enterprise A sends request to share Enterprise B's service, Message broker receives and validates the request, identifies required services from service registry by applying set of rules and delivers the necessary information regarding the identified services to communication handler. Communication handler calls integration bus to deliver the created service proxy to the requestor. Integration bus, a key component of SOA, supports asynchronous messaging, document exchange and above all provides powerful platform for connecting different applications together enabling seamless integration between components. Before delivering the service proxy to the requestor, it assesses the security issue by firing the trigger to the Functional analyzer.

Functionality analyzer analyzes Service Level Agreement (SLA) and policy defined between the two enterprises, identifies the list of constraints for integrating the service. Through this it scrutinizes the security gap between approved security policies and created service proxy and transmits the result to integration bus. Subsequently, integration bus handovers the proxy to the requestor. When requestor attempts to integrate the service, Property evaluation, heart of this model, validates integrating service with various constraints listed out by Functional analyzer to achieve the interoperability goals such as union, substitution, composition, finiteness, enhancement and configuration, etc.,. We will see the process of property evaluation detailly in next section. Evaluation metrics holds set of formulas to measure the activities and performance of service integration in order to achieve the interoperability goals efficiently. Business logic and rules are shared in such a way integration policy and interoperability goals are satisfied.

| Service Integration Request | B2B Communication System | Business Analyst |
|---|---|---|

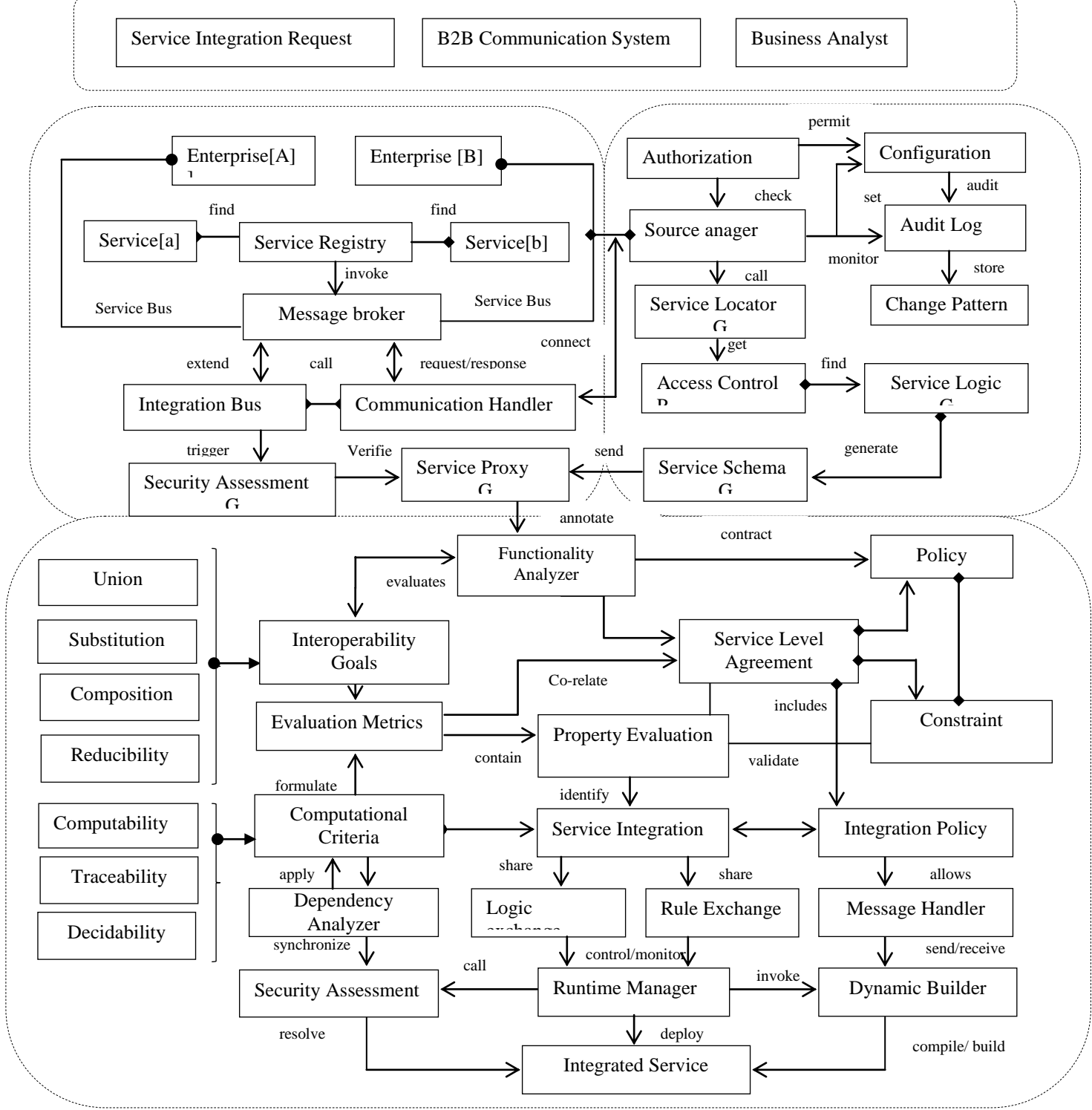

**Fig 1.** Detailed Architecture of Dynamic Web Service Integration

Message handler sends needed information about the service logic to the requestor. Work flow decider evaluates the performance of service integration through formulated metrics and sends the result to Exception handler. Runtime manager monitors the service logic while integrating with theirs, if at any case service integration violates the integration policy or deviates the interoperability goals, it calls exception handler. Exception handler handles and resolves the exception in such a way metric evaluated is also improved. Runtime manger invokes dynamic builder to build the newly integrated service dynamically and deploys the service in server. It monitors the service whenever changes have been done and redeploys dynamically. Source Manager monitors all these activities and adds necessary information to configuration and audit log.

## 4    Property Evaluation for Service Integration

### 4.1 Computability

Computability is an essential criterion in web service which determines whether the modified service is computable with in time limit.

Example The requirement is to create a service, e-payment to calculate total price for the list of purchased items and to transact the calculated amount. In the existing shopping application, we have billing service which computes total cost for the purchased items and transaction service in banking application transacts the amount. By integrating these two services, required new service e-payment can be developed. Here integration should be done in such a way that the processing time of the integrated service bounded within a time limit.

*logic1*

$BL_1$: public string billing(){
$BF_1$:   String username=username.get();
String password=password.get();
$DR_{f1}$:   String   sql="select   *   from   shopping   where   username="+username+"   and password="+password;
ResultSet rs=st.executeQuery(sql);
$CR_{r1}$:   if(rs.next()){
$BF_{r1}$:   double amount=calculateamount();
String accno=accountno.get();
$BF_{r2}$      String accno1=123456;
$BF_{f1}$:   String result="Amount to be paid="+amount;
$P_1$:   return result;
}}
*logic 2*
$BL_2$:  public string transact(){
$BF_{21}$   String accno=accno.get();
String accno1=accno1.get();
String amount=amount.get();
$BF_{22}$   String transid1=transid.set();
$DR_{f1}$   Statement st=con.createStatement();
ResultSet rs=st.executeQuery("select Balance from bank where Accountno="+ accno+'"');
$DR_{r1}$   double balance=rs.getDouble("Balance");
$CR_{r1}$ if( (balance-amount)>1000 ){
$DR_{rr1}$      st.executeUpdate("update   bank   set   balance=   balance-   "+amount+"   where Accountno="+accno+"";);
$DR_{rr2}$      st.executeUpdate(update   bank   set   balance=   balance+"+amount+"   where Accountno="+accno1+"");
$BF_{f1}$   String transid=" Amount"+amount+"transferred from"+accno+" to "+accno1;
$BF_{r2}$   String result= "Ur transaction id is "+transid1+" Ur transaction completed successfully";

P₂       return result;}

***Solution : Integrated logic***

$BL_1$ public string ebilling(){

$BF_{l1}$      String username=username.get();

String password=password.get();

$DR_{fl1:}$ String sql="select * from shopping where username="+username+" and password="+password;

ResultSet rs=st.executeQuery(sql);

$CR_{lfr1}$ if(rs.next()){

$BF_{lfrr1}$ double amount=calculateamount();

String accno=accountno.get();

String accno1=123456;

$BF_{lfrr1}$ transact(accno,amt,accno1);}

$BL_2$     public String transact(String accno, double amt, String accno1){

$BF_{l1}$      String transid1=transid.get();

$DR_{lf1}$ResultSet rs=st.executeQuery("select Balance from bank where Accountno='"+ accno+"'''");

$DR_{lfr1}$ double balance=rs.getDouble("Balance");

$CR_{lfrr1}$ if( (balance-amount)>1000 ){

$DR_{lfrrr1}$String sql="update bank set balance= balance- "+amount+" where Accountno='"+accno+"'";

st.executeUpdate(sql);

$DR_{lfrrrr1}$ sql="update bank set balance= balance+"+amount+" where Accountno='"+accno1+"'";

 st.executeUpdate(sql);

$P_{lfrrr1}$String transid=" Amount"+amount+"transferred from"+accno+" to "+accno1;

$P_{lfrrr2}$String result= "Ur transaction id is "+transid1+" Ur transaction completed successfully";}

***Logic Flow Diagram***

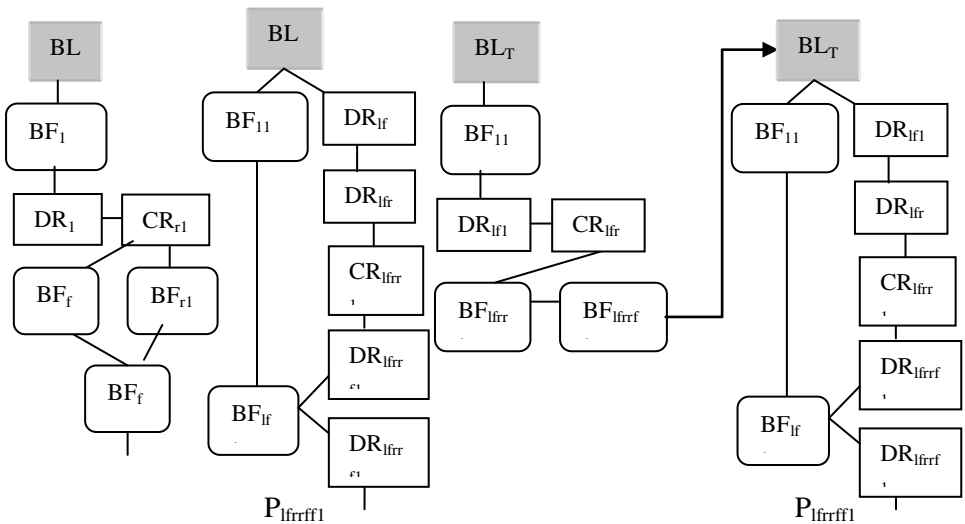

$BL_1 \rightarrow \{ BF_1\}$     $BL_2 \rightarrow \{[ BF_{lf1}, BF_{11}]\}$     $BL_{T1} \rightarrow \{ BF_{11}\}$

$BF_1 \rightarrow \{DR_{fl}\}$     $BF_{lf1} \rightarrow \{ DR_{lfr1}\}BF_{11} \rightarrow \{BF_{l1}\}$     $BF_{11} \rightarrow \{DR_{lf1}\}$

$DR_{fl} \rightarrow \{ CR_{r1}\}$     $DR_{lfr1} \rightarrow \{CR_{lfrr1}\}$     $DR_{lf1} \rightarrow \{ CR_{lfr1}\}$

$CR_{r1} \rightarrow \{ BF_{r1}\}$     $CR_{lfrr1} \rightarrow \{[DR_{lfrrf1}],[ BF_{l1}]\}$     $CR_{lfr1} \rightarrow \{ BF_{lfrr1}\}$

$BF_{r1} \rightarrow \{ BF_{fl}\}$     $DR_{lfrrf1} \rightarrow \{DR_{lfrrf1}\}$     $BF_{lfrr1} \rightarrow \{BF_{lfrrf1}\}$

$BF_{fl} \rightarrow \{P_1\}$     $DR_{lfrrf1} \rightarrow \{ P_{lfrrrf1}\}$     $BF_{lfrr1} \rightarrow \{ BL_{T2}\}$

                                                 $BL_{T2} \rightarrow \{[BF_{lf1}, BF_{11}]\}, BF_{11}= BF_{lfrr1}$

                                                 $BF_{lf1} \rightarrow \{ DR_{lf1}\}BF_{11} \rightarrow \{BF_{lf1}\}$

$$DR_{lfr1} \rightarrow \{CR_{lfr1}\}$$
$$CR_{lfrr1} \rightarrow \{[DR_{lfrrf1}],[\ _{BFlf1}]\}$$
$$DR_{lfrrf1} \rightarrow \{DR_{lfrrf1}\}$$
$$DR_{lfrrf1} \rightarrow \{\ P_{lfrrf1}\}$$

billing→{get} transaction→{get,set}
get→ {r:select} get→{ r:select}
r:select →{r:cmp} r:select→{r:cmp}
r:cmp →{compute} r:cmp→{r:update1,r:update2}
compute→{store} {r:update1, r:update2}→{store}
store→{return} store→{return}

billing→{get}
get→ {r:select}
r:select →{r:cmp}
r:cmp →{compute}
compute→{store}
store →{transaction}
transaction→{set,get}get=compute
get→{ r:select}
r:select→{r:cmp}
r:cmp→{[r:update1, r:update2]}
{r:update1, r:update2}→{store}
store→{return}

*BLPS of Logic 1*

```
1 ⊟ <service name="billing">
2  ⊟   <property>
3          <computability CF= BF1, BFf1,BFr1,BFr2 ,DR1,CRr1/>
4  ⊟   </property>
5  ⊟   <function name=get type="input"  index=BF1 >
6          <param1 name=username datatype='string' value=null/>
7          <param2 name=password datatype='string' value=null/>
8      </function>
9  ⊟   <rule index=DR1 name="select" type="data manipulation" dbname="shopping" >
10 ⊟      <conditions setStatus=true>
11            <condition1 Lvar= db.username expr= &eq rvar=$username />
12            <condition2 lvar= db.password expr=&eq rvar=$password />
13         </conditions>
14         <retrieve param1=accno />
15      </rule>
16 ⊟   <rule index=CRr1  name="if" type="conditional" condition:status=true>
17         <function index=BFr1 name=compute '' action='' call' target-function=compute() return-type="double"  store-result=amount />
18         <function index=BFr2 name=set type=set >
19            <param name=accno1 value='123456'/>
20         </function>
21         <function index=BFf1 name=assign type=output> <param name=result returntype=double value=amount/>
22      </rule>
23    </service>
```

*BLPS of Logic 2*

```
1 ⊟ <service name="transact" >
2  ⊟   <property>
3          <computability CF="BF21,DRf1,CRr1,DRrr1, DRrr2"/>
4      </property>
5  ⊟   <function name="get" type="input"  index=BF21>
6          <param1 name=accno datatype="string" value=null/>
7          <param2 name=accno1 datatype="string" value=null/>
8          <param2 name=amount datatype="double" value=null/>
9      </function>
10 ⊟   <rule index=DRf1 name="select" type="data manipulation" dbname="bank" >
11         <condition1 Lvar= db.accountno expr= &eq rvar=$accountno />
12         <retrieve param1=balance/>
13 ⊟      <rule index=CRr1  name="if" type="conditional">
14            <condition lvar=$balance-$amount expr=> rvar=1000>
15            <rule index=DRrr1 name="update" type="data manipulation" dbname="bank" >
16               <condition Lvar= db.accountno expr= &eq rvar=$accno />
17               <assign Lvar=balance expr=&eq rvar=($balance-$amount) />
18      </rule>
19 ⊟   <rule index=DRrr2 name="update" type="data manipulation" dbname="bank" >
20            <condition Lvar= db.accountno expr= &eq rvar=$accno1 />
21            <assign Lvar=balance expr=&eq rvar=($balance+$amount) />
22      </rule>
23         <function index=BFf1 name=assign type=output> <param name=transid returntype=double value=transstat/>
24      </rule>
25    </service>
```

*Integrated Service*

```xml
1  <service name="e-billing">
2    <property>
3      <computable CF=BF1, BFf1,BFr1,BFr2,BF21,DR1,CRr1, DRf1,CRr1,DRrr1, DRrr2/>
4    </property>
5    <service name="billing">
6      <function name="get type="input"  index=BF1>
7        <param1 name=username datatype="string" value=null>
8        <param2 name=password datatype="string" value=null>
9      <rule index=DR1 name="select" type="Data Manipulation" dbname="shopping" >
10       <conditions  set(status=true>
11         <condition1 Lvar= db.username exp= &eq rvar=$username
12         <condition2 Lvar= db.password exp= &eq rvar=$password
13       </conditions>
14       <retrieve param1=accno/>
15     </rule>
16     <rule index=CRr1  name="if" type="conditional" condition:status=true>
17       <function index=BFr1 name=compute " action='call' target-function=compute() return-type="double" store-result=amount />
18       <function index=BFr2 name=set type=set >
19         <param name=accno1 value= 123456'/>
20       </function>
21       <function index=BFf1 name=call type=invoke target-service=transact>
22         <arg  name=accno returntype=string>
23         <arg  name=accno1 returntype=string>
24         <arg  name=amount returntype=double>
25       </function>
26     </rule>
27   </service>
28 </service>
29 <service name=transact>
30   <function name=receive >
31     <arg  name=accno datatype=string>
32     <arg  name=accno1 returntype=string>
33     <arg  name=amount datatype=double>
34   </function>
35   <function name=get type="input"  index=BF1>
36     <param1 name=accno datatype="string" value=$accno>
37     <param2 name=accno1 datatype="string" value=$accno1>
38     <param2 name=amount datatype="double" value=$amount>
39   </function>
40   <rule index=DR1 name="select" type="data manipulation" dbname="bank" >
41     <condition1 Lvar= db.accountno exp= &eq rvar=$accountno />
42     <retrieve param1=balance>
43   </rule>
44   <rule index=CRr1 name="if" type="conditional">
45     <condition lvar=$balance-$amount exp=> rvar=1000>
46       <rule index=DR1 name="update" type="data manipulation" dbname="bank" >
47         <condition Lvar= db.accountno exp= &eq rvar=$accno />
48         <assign Lvar=balance exp=&eq rvar=($balance-$amount)>
49       </rule>
50     </condition>
51   </rule>
52   <rule index=DR1 name="update" type="data manipulation" dbname="bank" >
53     <condition Lvar= db.accountno exp= &eq rvar=$accno />
54     <assign Lvar=balance exp=&eq rvar=($balance-$amount)>
55   </rule>
56   <function index=BFf1 name=assign type=output>
57     <param name=transid returntype=double value=transstat/>
58   </function>
59 </service>
```

## 4.2 Traceability

Traceability in general is 'ability to chronologically interrelate the uniquely identifiable entities in a way that matters'. It verifies the flow, assesses the risk, checks completeness and helps to improve the quality by tracing each and every step of the service.

Example: In the previous case, integrated service might fail due to transaction failure or erroneous calculation of price. So it is necessary to trace the service and verify the transaction status at the end of every transaction. Transaction id gives necessary information of that transaction such as credit, debit, time, etc. So it is enough to trace the transaction id to verify the whole service.

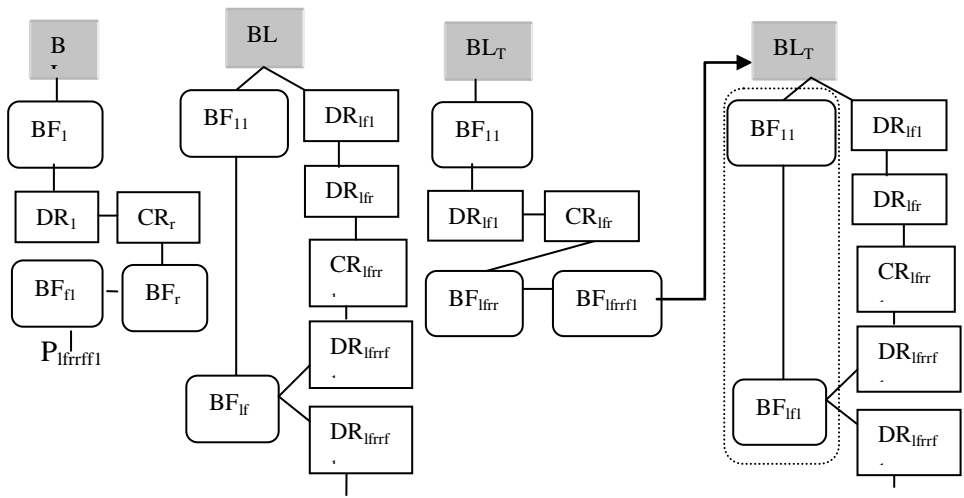

*Input 1*

```
1  <service name="billing">
2    <property>
3        <computability CF= BF1, BFf1,BFr1,BFr2 ,DR1,CRr1/>
4    </property>
5        <function name=get type="input"  index=BF1>
6            <param1 name=username datatype="string" value=null/>
7            <param2 name=password datatype="string" value=null/>
8        </function>
9    <rule index=DR1 name="select" type="data manipulation" dbname="shopping" >
10          <conditions setStatus=true>
11              <condition1 Lvar= db.username expr= &eq rvar=$username />
12              <condition2  lvar= db.password expr= &eq rvar=$password />
13          </conditions>
14          <retrieve param1=accno/>
15      </rule>
16   <rule index=CRr1  name="if" type="conditional" condition:status=true>
17          <function index=BFr1 name=compute " action='call' target-function=compute() return-type="double"  store-result=amount />
18          <function index=BFr1 name=set type=set >
19              <param name=accno1 value="123456"/>
20          </function>
21          <function index=BFf1 name=assign type=output <param name=result returntype=double value=amount/>
22      </rule>
23   </service>
```

*Input 2*

```
1  <service name="transact" >
2    <property>
3        <computability CF="BF21,DRf1,CRr1,DRrr1, DRrr2"/>
4    </property>
5        <function name=get type="input"  index=BF21>
6            <param1 name=accno datatype="string" value=null/>
7            <param2 name=accno1 datatype="string" value=null/>
8            <param2 name=amount datatype="double" value=null/>
9        </function>
10   <rule index=DRrf1 name="select" type="data manipulation" dbname="bank" >
11          <condition1 Lvar= db.accountno expr= &eq rvar=$accountno />
12          <retrieve param1=balance/>
13          <rule index=CRr1  name="if" type="conditional" >
14              <condition lvar=$balance-$amount expr=> rvar=1000>
15              <rule index=DRrr1 name="update" type="data manipulation" dbname="bank" >
16              <condition Lvar= db.accountno expr= &eq rvar=$accno />
17              <assign Lvar=balance expr=&eq rvar=($balance-$amount) />
18          </rule>
19   <rule index=DRrr2 name="update" type="data manipulation" dbname="bank" >
20          <condition Lvar= db.accountno expr= &eq rvar=$accno1 />
21          <assign Lvar=balance expr=&eq rvar=($balance+$amount) />
22      </rule>
23      <function index=BFf1 name=assign type=output <param name=transid returntype=double value=transstat/>
24      </rule>
25   </service>
```

*Service Integration Solution (BLP Schema)*

```xml
1  <service name="e-billing">
2    <property>
3      <traceability TF=" BF1, BFf1" NTF="BFf1,BFr1,BFr2 , DR1,CRr1, DRf1,CRr1,DRrr1, DRrr2"/>
4    </property>
5    <service name="billing">
6      <function name="get" type="input" index=BF1>
7        <param1 name=username datatype='string' value=null />
8        <param2 name=password datatype='string' value=null />
9      </function>
10     <rule index=DR1 name="select" type="Data Manipulation" dbname="shopping" >
11       <conditions setStatus=true>
12         <condition1 Lvar= db.username expr= &eq rvar=$username
13         <condition2 Lvar= db.password expr= &eq rvar=$password
14         <retrieve param1=amount>
15     </rule>
16     <rule index=CRr1 type="conditional" condition:status=true>
17       <function index=BFr1 name="compute" action="call" target-function=compute() return-type="double"  store-result=amount />
18       <function index=BFr2 name= set type=set >
19         <param name=accno1 value="123456"/>
20       </functions>
21     <function index=BFf1 name=call type=invoke target-service=transact>
22         <arg  name=accno returntype=string>
23         <arg  name=accno1 returntype=string>
24         <arg  name=amount returntype=double>
25     </function>
26   </rule>
27   </service>
28   <service name=transact>
29     <function name=receive>
30       <arg  name=accno datatype=string>
31       <arg  name=accno1 datatype=string>
32       <arg  name=amount datatype=double>
33     </function>
34     <function name=get type="input"  index=BF1>
35       <param1 name=accno datatype='string'  value=$accno
36       <param2 name=accno1 datatype='string'  value=$accno1>
37       <param2 name=amount datatype='double'  value=$amount>
38     </function>
39     <rule index=DR1 name="select" type="data manipulation" dbname="bank" >
40       <conditions setStatus=true>
41         <condition1 Lvar= db.accountno expr= &eq rvar=$accountno
42         <retrieve param1=balance>
43     </rule>
44     <rule index=CRr1 name="if" type="conditional">
45       <condition lvar=$balance+$amount expr=> rvar=1000</>
46       <rule index=DR1 name="update" type="data manipulation" dbname="bank" >
47         <condition Lvar= db.accountno expr= &eq rvar=$accno
48         <assign Lvar=balance expr=&eq rvar=($balance-$amount)>
49     </rule>
50     <rule index=DR1 name="update" type="data manipulation" dbname="bank" >
51       <condition Lvar= db.accountno expr= &eq rvar=$accno1
52       <assign Lvar= balance expr=&eq rvar=($balance+$amount)>
53     </rule>
54     <function index=BFf1 name=assign type=output>
55       <param name=transid returntype=double value=transstat>
56     </function>
57   </service>
58 </service>
```

## 4.3 Accessibility

Definition: Accessibility defines the extent to which one service can access the other service's logic.

Example:
The requirement is to create a new service, e-payment to calculate total price for the list of purchased items and to transact the calculated amount. In the existing shopping application, we have billing service which computes total cost for the purchased items and transaction service in banking application transacts the amount. By integrating these two services, required new service e-payment can be developed. Here integration should be done in such a way transaction service could access only the information returned by billing service, it should not view customer's credential information.

**logic1:**

$BL_1$: public string billing(){
$BF_{l1}$:     String username=username.get();
               String password=password.get();
$DR_{fl1}$: String sql="select * from shopping where username="+username+" and
           password="+password;
           ResultSet rs=st.executeQuery(sql);
$CR_{lfr1}$:   if(rs.next()){
$BF_{lfr1}$: double amount=calculateamount();
             String accno=accountno.get();
             String accno1=123456;
$BF_{lfrf1}$: String result="Amount to be paid="+amount;
$P_{lfrff1}$:     return result;
             }}

**logic 2;**

$BL_2$: public string transact(){
$BF_{l1}$    String accno=accno.get();
             String accno1=accno1.get();

String amount=amount.get();
                String transid1=transid.create();
DR$_{lf1}$ Statement st=con.createStatement();
                ResultSet rs=st.executeQuery("select Balance from bank where Accountno='"+
                accno+"'");
DR$_{lfr1}$  double balance=rs.getDouble("Balance");
CR$_{lfrr1}$ if( (balance-amount)>1000 ){
DR$_{lfrrr1}$ st.executeUpdate("update bank set balance= balance- "+amount+" where
Accountno='"+accno+"'";);
 DR$_{lfrrr2}$ st.executeUpdate(update bank set balance= balance+"+amount+" where
Accountno='"+accno1+"'");
BF$_{lf1}$    String transid=" Amount"+amount+"transferred from"+accno+" to "+accno1;
BF$_{lfrrr2}$    String result= "Ur transaction id is "+transid1+" Ur transaction completed
                successfully";
P$_{lfrrrf1}$    return result;
}

**Integrated logic:**
BL$_1$ public string ebilling(){
BF$_{l1}$     String username=username.get();
                String password=password.get();

DR$_{fl1}$: String sql="select * from shopping where username='"+username+" and
                password='"+password;
                ResultSet rs=st.executeQuery(sql);
CR$_{lfr1}$ if(rs.next()){
BF$_{lfrr1}$ double amount=calculateamount();
                String accno=accountno.get();
                String accno1=123456;
BF$_{lfrr1}$ transact(accno,amt,accno1);
}
BL$_2$    public String transact(String accno, double amt, String accno1){
BF$_{l1}$      String transid1=transid.get();
DR$_{lf1}$ResultSet rs=st.executeQuery("select Balance from bank where Accountno='"+
                accno+"'");
DR$_{lfr1}$ double balance=rs.getDouble("Balance");
CR$_{lfrr1}$ if( (balance-amount)>1000 ){
DR$_{lfrrr1}$String sql="update bank set balance= balance- "+amount+" where
                Accountno='"+accno+"'";
                st.executeUpdate(sql);
DR$_{lfrrrr1}$ sql="update bank set balance= balance+"+amount+" where Accountno='"+accno1+"'";
                st.executeUpdate(sql);
P$_{lfrrr1}$String transid=" Amount"+amount+"transferred from"+accno+" to "+accno1;
P$_{lfrr2}$String result= "Ur transaction id is "+transid1+" Ur transaction completed successfully";}

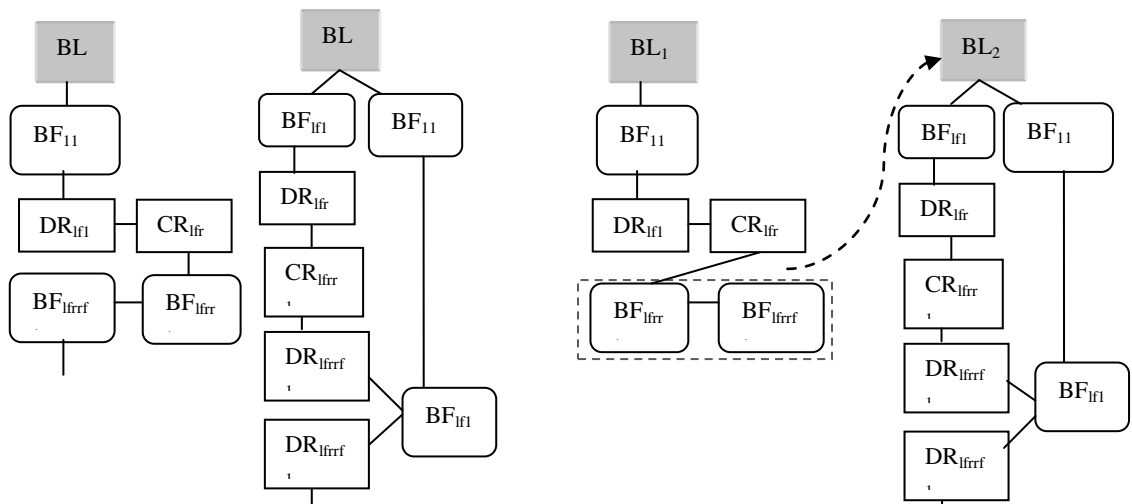

P$_{lfrrf1}$

P$_{lfffrrf1}$                                                            P$_{lffrrf1}$

BL$_1$→{ BF$_{11}$}          BL$_2$→{[ BF$_{lf1}$,BF$_{11}$]}          BL$_1$→{ BF$_{11}$}
BF$_{11}$→{DR$_{lf1}$}       BF$_{lf1}$→{ DR$_{lfr1}$}BF$_{11}$→{BF$_{lf1}$}   BF$_{11}$→{DR$_{lf1}$}
DR$_{lf1}$→{ CR$_{lfr1}$}    DR$_{lfr1}$→ {CR$_{lfrr1}$}               DR$_{lf1}$→{ CR$_{lfr1}$}
CR$_{lfr1}$→{ BF$_{lfrr1}$}  CR$_{lfrr1}$→ {[DR$_{lfrrf1}$],[ $_{BFlf1}$]}   CR$_{lfr1}$→{ BF$_{lfrr1}$}
BF$_{lfrr1}$→{ BF$_{lfrrf1}$}  DR$_{lfrrf1}$→ {DR$_{lfrrf1}$}          BF$_{lfrr1}$→{BF$_{lfrrf1}$}
BF$_{lfrrf1}$→{P$_{lfrrf1}$}   DR$_{lfrrf1}$→ { P$_{lfrrf1}$}          BL$_2$→{[ BF$_{lf1}$,BF$_{11}$]}
                                                                    BF$_{lf1}$→{ DR$_{lfr1}$}BF$_{11}$→{BF$_{lf1}$}
                                                                    DR$_{lfr1}$→ {CR$_{lfrr1}$}
                                                                    CR$_{lfrr1}$→ {[DR$_{lfrrf1}$],[ $_{BFlf1}$]}
                                                                    DR$_{lfrrf1}$→ {DR$_{lfrrf1}$}
                                                                    DR$_{lfrrf1}$→ { P$_{lfrrf1}$}

billing→{get}               transaction→{get,set}                    billing→{get}
get→ { r:select}            get→{ r:select}                          get→ {r:select}
r:select→{r:cmp}            r:select→{r:cmp}                         r:select→{r:cmp}
r:cmp →{compute}           r:cmp→{r:update1, r:update2}             r:cmp →{compute}
compute→{store}            {r:update1, r:update2}→{store}           compute→{store}
store→{return}             store→{return}                           store →{transaction}
                                                                    transaction→{set,get}get=compute
                                                                    get→{ r:select}
                                                                    r:select→{r:cmp}
                                                                    r:cmp→{[r:update1, r:update2]}
                                                                    {r:update1, r:update2}→{store}
                                                                    store→{return}

Logic 1

```
1 ⊟ <service name="billing">
2    <property>
3       <Accessibility AF= "CRr1, BFf1" NAF="BF1,BFr1,BFr2 ,DR1"/>
4    </property>
5 ⊟  <function name=get type="input"  index=BF1>
6       <param1 name=username datatype='string' value=null/>
7       <param2 name=password datatype='string' value=null/>
8    </function>
9 ⊟  <rule index=DR1 name="select' type="data manipulation" dbname="shopping" >
10 ⊟     <conditions setStatus=true>
11         <condition1 Lvar= db.username expr= &eq rvar=$username />
12         <condition2  lvar= db.password expr=&eq rvar=$password />
13      </conditions>
14      <retrieve param1=accno/>
15    </rule>
16 ⊟  <rule index=CRr1  name="if" type='conditional' condition:status=true>
17      <function index=BFr1 name=compute '' action='call' target-function=compute() return-type="double"  store-result=amount />
18 ⊟     <function index=BFr2 name=set type=set >
19         <param name=accno1 value='123456'/>
20      </function>
21 ⊟     <function index=BFf1 name=assign type=output>
22         <param name=result returntype=double value=amount/>
23      </function>
24    </rule>
25  </service>
26
```

## Logic 2

```
1   <service name="transact" >
2       <property>
3           <Accessibility AF="BF21" NAF="BF21,DRf1,CRr1,DRrr1, DRrr2"/>
4       </property>
5       <function name=get type="input"   index=BF21>
6           <param name=accno datatype='string' value=null/>
7           <param2 name=accno1 datatype='string' value=null/>
8           <param2 name=amount datatype='double' value=null/>
9       </function>
10      <rule index=DRf1 name="select' type="data manipulation" dbname="bank" >
11          <condition1 Lvar= db.accountno expr= &eq rvar=$accountno />
12          <retrieve param1=balance/>
13          <rule index=CRr1  name="if" type="conditional'>
14              <condition lvar=$balance-$amount expr=> rvar=1000/>
15              <rule index=DRrr1 name="update' type="data manipulation"
                    dbname="bank" >
16              <condition Lvar= db.accountno expr= &eq rvar=$accno />
17              <assign Lvar=balance expr=&eq rvar=($balance-$amount) />
18      </rule>
19      <rule index=DRrr2 name="update' type="data manipulation" dbname="bank" >
20          <condition Lvar= db.accountno expr= &eq rvar=$accno1 />
21          <assign Lvar=balance expr=&eq rvar=($balance+$amount) />
22      </rule>
23      <function index=BFf1 name=assign type=output>
24          <param name=transid returntype=string value=transstat/>
25      </function>
26      </rule>
27  </service>
28
```

```
1  <service name="e-billing">
2      <property>
3          <computable CF=CRr1,BFf1 NCF=BF1,BFr1,BFr2,BF21,DR1,DRf1,CRr1,DRrr1,DRrr2/>
4      </property>
5      <service name="billing">
6          <function name=get type="input"  index=BF1>
7              <param1 name=username datatype='string' value=null/>
8              <param2 name=password datatype='string' value=null/>
9          </function>
10         <rule index=DR1 name='select' type="Data Manipulation" dbname="shopping" >
11             <conditions setStatus=true>
12                 <condition1 Lvar= db.username expr= &eq rvar=$username />
13                 <condition2 lvar= db.password expr=&eq rvar=$password />
14             </conditions.
15             <retrieve param1=accno />
16         </rule>
17         <rule index=CRr1  name="if" type="conditional' condition:status=true>
18             <function index=BFr1 name=compute " action='call' target-function=compute() return-type="double"  store-result=amount />
19             <function index=BFr2 name=set type=set >
20                 <param name=accno1 value='123456'/>
21             </function>
22             <function index=BFf1 name=call type=invoke target-service=transact>
23                 <arg  name=accno returntype=string>
24                 <arg  name=accno1 returntype=string>
25                 <arg  name=amount returntype=double>
26             </function>
27         </rule>
28     </service>
29     <service name=transact>
30     <function name=receive >
31         <arg  name=accno datatype=string>
32         <arg  name=accno1 returntype=string>
33         <arg  name=amount datatype=double>
34     </function>
35     <function name=get type="input"  index=BF1>
36         <param1 name=accno datatype='string' value=$accno/>
37         <param2 name=accno1 datatype='string' value=$accno1/>
38         <param3 name=amount datatype='double' value=$amount/>
39     </function>
40     <rule index=DR1 name="select' type="data manipulation" dbname="bank">
41         <condition1 Lvar= db.accountno expr= &eq rvar=$accountno />
42         <retrieve param1=balance />
43     </rule>
44     <rule index=CRr1  name="if" type="conditional'>
45         <condition lvar=$balance+$amount expr=> rvar=1000>
46             <rule index=DR1 name='update' type="data manipulation" dbname="bank" >
47                 <condition Lvar= db.accountno expr= &eq rvar=$accno />
48                 <assign Lvar=balance expr=&eq rvar=(&balance-$amount) />
49             </rule>
50         </condition>
51     </rule>
52     <rule index=DR1 name='update' type="data manipulation" dbname="bank">
53         <condition Lvar= db.accountno expr= &eq rvar=$accno1 />
54         <assign Lvar=balance expr=&eq rvar=(&balance+$amount) />
55     </rule>
56     <function index=BFf1 name=assign type=output>
57         <param name=transid returntype=double value=transstat/>
58     </function>
59 </service>
60 </service>
```

Here BL$_2$ can access only the highlighted part of service BL$_1$.

## 5. IMPLEMENTATION METHODOLOGY

The web service online payment system is developed by integrating existing billing service and transaction service in banking application as discussed above. Computability and traceability properties are verified as discussed in last section. BPEL diagram of newly developed service is depicted in Fig 2.

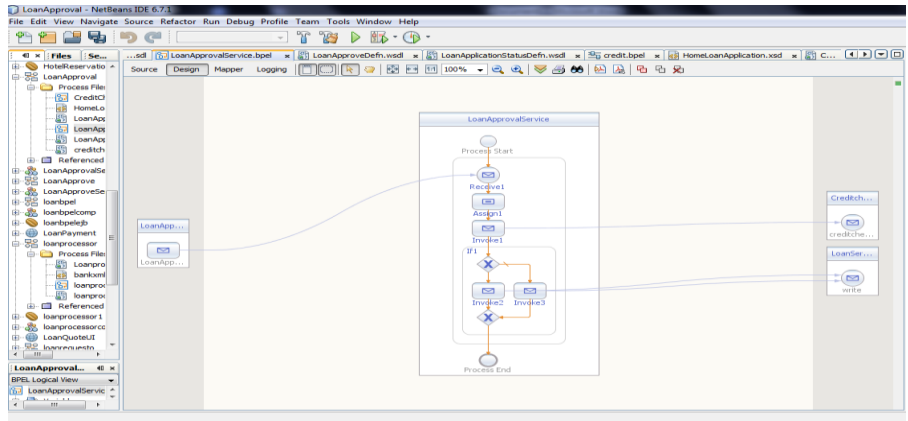

**Fig 2.** Service Integration using BLP schema

## 6. CONCLUSION

The proposed model provides a powerful platform to share service logic dynamically and securely in such way interoperability between the services is managed. This paper evaluates the services to be integrated with properties such as computability, traceability and accessibility and integrates in efficient way. Also, this model progressively monitors the changes made in the source code and points out whether the changes made affect the computability and traceability criteria's of the web services. Examples given in this paper explains how properties are evaluated for various situations. This would be a standard platform for service providers to share their resources dynamically and securely.

## References

1. Zhuoren Jiang, Yan Chen and Ming Yang, "A research on multi-layer structure for dynamic service integration", IEEE international conference, 2010.

2. W.J. Yan, P.S. Tan and E.W. Lee," A Web Services-enabled B2B Integration Approach for SMEs", IEEE international Conference on Industrial Informatics, July 13-16, 2008.

3. Liyi Zhang and Si Zhou, "A Semantic Service Oriented Architecture for Enterprise Application Integration", Second International Symposium on Electronic Commerce and Security, 2009.

4. Thomas Haselwanter, Paavo Kotinurmi, Matthew Moran, Tomas Vitvar, and Maciej Zaremba, "WSMX: A Semantic Service Oriented Middleware for B2B Integration", available at http://www.vitvar.com/tomas/!publications/icsoc2006-WSMX.pdf.

5. Jianwei Yin, Hanwei Chen, Shuiguang Deng and Zhaohui Wu, "A Dependable ESB framework for Service Integration", IEEE Internet Computing, 2009.

6. Gulnoza Ziyaeva, Eunmi Choi and Dugki Min, "Content-Based Intelligent Routing and Message Processing in Enterprise Service Bus", International Conference on Convergence and Hybrid Information Technology, 2008.

7. Soo Ho Chang, Jeong Seop Bae, Won Young Jeon, Hyun Jung La, and Soo Dong Kim, "A Practical Framework for Dynamic Composition on Enterprise Service Bus", IEEE international conference on Service Computing, 2007.

8.  Liu Ying and Wang Li, "An Intelligent Service Composer for Business-level Service Composition", Nineth international conference on Enterprise Computing, E-Commerce and E-Services, 2007.

9.  http://en.wikipedia.org/wiki/Electronic_Data_Interchange.

10. Rossatanet, "http://www.rosettanet.org".

11. Jing Wang and Yeong-Tae Song, "Architectures Supporting RosettaNet", Proceedings of the Fourth International Conference on Software Engineering Research,2006.

12. ebxml," http://www.ebxml.org ".